\documentstyle[prl,aps,multicol,epsf]{revtex}
\begin{document}
\draft
\title{Simple Ginzburg-Landau Theory for Vortices in a
Crystal Lattice}
\author{
Joonhyun Yeo and M. A. Moore} 
\address{Department of Physics, University of Manchester,
Manchester, M13 9PL, United Kingdom.}
\date{\today}
\maketitle
\begin{abstract}
We study the Ginzburg-Landau model with a nonlocal quartic term
as a simple phenomenological model for superconductors in the presence of
coupling between the vortex lattice and the underlying crystal 
lattice. In mean-field theory, our model is consistent with
a general oblique vortex lattice ranging from a triangular
lattice to a square lattice. This simple formulation
enables us to study the effect of thermal fluctuations
in the vortex liquid regime. We calculate the structure
factor of the vortex liquid
nonperturbatively and find Bragg-like peaks with four-fold
symmetry appearing in the structure factor even though there is
only a short-range crystalline order.
\end{abstract}
\pacs{PACS numbers: 74.20.De, 74.60.Ge}
\begin{multicols}{2}
\narrowtext

It is of great interest to study vortex lattice structure
and correlations in superconductors 
in the presence of coupling between the vortex lattice and 
the underlying crystal lattice.
Various experimental probes including 
neutron diffraction \cite{neutron},
Bitter decoration \cite{lowh1},
and scanning tunneling microscopy \cite{stm,new}
have been used to reveal a range of vortex lattice structures
from the usual triangular lattice to a general oblique lattice
and a square lattice oriented along a specific direction of 
the crystal axis. An important feature of these structures
is the emergence of the four-fold symmetry representing the 
symmetry of the underlying crystal lattice. This effect of the
crystal lattice on the vortex lattice is found to be dependent 
upon the external field in such a way that a triangular vortex
lattice is observed at low fields, while at higher fields 
a square lattice is observed \cite{new}.

In order to study the vortex lattice structure, one usually
uses a Ginzburg-Landau (GL)  phenomenological theory.
Since the usual GL theory is rotationally invariant, one
needs additional terms that break this symmetry to 
account for the appearance of the four-fold symmetric vortex
lattice structure. The conventional way to include this effect
is to introduce  terms quadratic in the order parameter in
the GL free energy with fourth order derivatives \cite{new,aff,franz}.
 The observed four-fold symmetric vortex
lattice structure can be explained within these formalisms. 
However, the equations involved in these theories are very 
difficult to handle even at the linearized level where 
 one has to resort to  approximate or  numerical methods. 
 
In this paper we propose a much simpler phenomenological model
for vortices in a crystal lattice. Our model is the usual GL theory for a
one-component complex order parameter, except that the term quartic in the
order parameter, $\Psi ({\bf r})$ is {\em nonlocal}.
We consider the situation where the effect of the underlying crystal
on the order parameter symmetry can be summarized into
an appropriate form for the nonlocal interaction potential. 
The vortex lattice structure is determined within our model 
without much calculational effort.
We find that the four-fold symmetric vortex lattice structure 
can be modelled with
the appropriate choice of a minimal number of parameters describing
the nonlocal quartic interaction. The a-b plane anisotropy found in
 high-temperature superconductors 
is also incorporated in our model
in the usual way through second order gradient terms.   
We believe that this model captures the same physics as in the higher
derivative approaches of Ref.~\onlinecite{new,aff,franz},
but is just much simpler to handle calculationally.

One feature of our simple formulation
is that it allows a study of the effect of thermal fluctuations
in the vortex liquid regime. This would be quite
impossible in the conventional formulations.
We apply the nonperturbative method
developed by us in Ref.~\onlinecite{ym} to
calculate the structure factor of
the vortex liquid, which is measured in neutron
scattering experiments. One of the main results of the present work is
that one can observe in the structure factor
the ring patterns expected for a liquid state 
but broken up into Bragg-like peaks.
This suggests that even in the vortex liquid state where there is 
 a short-range crystalline order, 
a weak four-fold symmetric coupling to the 
underlying crystal may produce the 
spots observed in 
neutron scattering experiments and usually attributed to the 
formation of the vortex crystal state.

The model we study in this paper
is based on the GL free energy with
a nonlocal quartic interaction
for a two-dimensional superconductor in a magnetic
field ${\bf B}={\bf \nabla}\times {\bf A}$:
\begin{eqnarray}
F[\Psi ]&=&\int d^2{\bf r}\big(\frac{\hbar^2}{2m_x}|
D_x\Psi |^2 +\frac{\hbar^2}{2m_y}|
D_y\Psi |^2+\alpha |\Psi ({\bf r}) |^2\big) \nonumber \\
&&+\frac{\beta}{2}\int d^2 {\bf r}_1 d^2 {\bf r}_2
|\Psi ({\bf r}_1)|^2 g({\bf r}_1-{\bf r}_2) |\Psi
({\bf r}_2)|^2  ,
\label{eq1}
\end{eqnarray}
where $\alpha,\beta$ are phenomenological parameters,
$m_x$ the effective mass in the x-direction, $m_y$ in 
the y-direction,
and ${\bf D}=-i{\bf \nabla}-(e^*/\hbar c){\bf A}$.
The a-b plane anisotropy is represented by the ratio of
the effective masses. For later use, we define
$\sigma\equiv (m_x/m_y)^{1/4}$. The only difference between
our model and the usual
GL theory is the nonlocal quartic interaction term 
represented here by a general function   
$g({\bf r})=g(x,y)$, which is equal to the delta-function
$\delta^{(2)}({\bf r})$ for the usual local GL theory. 

One reason we study the two-dimensional 
version of the model in this paper is that
one can easily apply the nonperturbative method in Ref.~\onlinecite{ym} 
to calculate the vortex liquid structure factor. But, more 
importantly, as noted by one of us \cite{mike},
the phase correlation length 
parallel to the field direction
in a bulk superconductor grows exponentially as one approaches 
zero temperature. 
When this length scale becomes
comparable to or larger than the sample size
in the low temperature regime, 
the system effectively
behaves as a two-dimensional thin film with phase coherence
across the sample and a very low effective 
temperature \cite{mike}. We note that 
most of the experiments mentioned
earlier are performed in this regime, and we expect that they can be 
described by the present model when the parameter $\alpha$ is set to
large negative values, {\it i.e.} low temperatures.  

A nonlocal quartic interaction as in (\ref{eq1})
appeared  in the renormalization group study \cite{bnt} 
of this system near its
upper critical dimension. Even if one starts from 
a local theory,  renormalization always drives 
the quartic term into an effective nonlocal one. 
It is not our aim here 
to derive an explicit form of $g({\bf r})$ 
from a microscopic theory.
Instead, we take (\ref{eq1}) as our starting point for the
phenomenological description of superconductors in the 
presence of an interaction between the vortex lattice and the crystal 
lattice, and show that the variety of vortex lattice structures 
observed in experiments
can be explained using a very simple form of $g({\bf r})$.    

The main ingredient one has to incorporate into the phenomenological
construction of $g({\bf r})$ is the presumed four-fold symmetry of the
underlying crystal. (Generalizations to other crystal symmetries are of course
possible). In the present work, we construct $g({\bf r})$ such that it contains
a term with explicit four-fold symmetry in addition to a rotationally symmetric
term.
Thus, we take for the Fourier transform 
$\widetilde{g}({\bf k})\equiv \int d^2 {\bf r} g({\bf r})
\exp(i{\bf k}\cdot{\bf r})$,  
\begin{equation}
\widetilde{g}({\bf k})=\exp\big[ -C (\frac{k}{\mu})^4\big(
1-\varepsilon\cos 4(\theta+\theta_0)\big)\big], 
\label{gk:def} 
\end{equation}
where ${\bf k}=(k\cos\theta,k\sin\theta)$ and
$\mu=\sqrt{e^* B/\hbar c}$ is the inverse magnetic
length. In particular, $\widetilde{g}({\bf k})=
\exp[-C\{ (1-\varepsilon)(k^4_x+k^4_y)+2(1+3\varepsilon)
k^2_x k^2_y\}/\mu^4]$ for $\theta_0=0$. In (\ref{gk:def}), we 
introduced three parameters $C$, $\varepsilon$
and $\theta_0$ in such a way that
the overall constant $C$ controls the
strength of the interaction between the vortex lattice and the
underlying crystal
(note that when $C\rightarrow 0$, a local GL theory
is recovered), and the
dimensionless parameter $\varepsilon$ $(0\le\varepsilon <1)$ represents
the strength of the four-fold symmetric interaction
compared to the rotationally symmetric one. As will be discussed
later, $\theta_0$ controls the orientation of the vortex lattice 
with respect to the underlying crystal. 
This choice for $\widetilde{g}$ is certainly not unique.
One could use various other forms with a four-fold symmetric 
term as long as they are positive definite (to ensure the stability of
GL theory) and they do not introduce any unphysical non-analyticity
(Note that in order 
to ensure the analyticity of $\widetilde{g}({\bf k})$ near ${\bf k}=0$
with the $\cos 4\theta$ term, 
we need at least $k^4$ terms in (\ref{gk:def})).
But, in the presence of the growing length scale mentioned earlier,
we might expect that universality applies and the present form
of the nonlocal quartic interaction does not alter the physical
results. Note that 
we measure the wave-vectors with respect to the inverse magnetic 
length. But this is just for convenience in later calculations.
A more natural length scale for the nonlocal kernel will be 
the lattice spacing of the crystal lattice, $l_0$.
Therefore, the dimensionless parameter $C$ which appears in 
(\ref{gk:def}) will depend on 
the ratio of two length scales,
{\it i.e.}
$C\sim (l_0\mu)^4$. 
As the magnetic field increases, the inter-vortex spacing,
$\mu^{-1}$, gets smaller, so the nonlocal interaction term
becomes more important. This qualitative feature 
of our model is consistent with 
experimental findings
where the four-fold symmetric vortex lattice
structure is observed only in the high field regime.

It is convenient to map the free energy functional of Eq.~(\ref{eq1})
into a form with isotropic gradient terms using the 
following transformations:
${\bf a} ({\bf r}^\prime)=(a_x ({\bf r}^\prime), 
a_y ({\bf r}^\prime))\equiv
(\sigma^{-1} A_x ({\bf r}), \sigma A_y ({\bf r}))$, and
$h ({\bf r}^\prime)\equiv g({\bf r})$, where
${\bf r}^\prime =(x^\prime,y^\prime)\equiv 
(\sigma x, \sigma^{-1} y)$, or the Fourier transform
$\widetilde{h}({\bf k}^\prime)=
\widetilde{g} (\sigma k^\prime_x , \sigma^{-1} k^\prime_y)$.
Then Eq.~(\ref{eq1}) becomes in
the new order parameter $
\psi ({\bf r}^\prime)\equiv\Psi ({\bf r})=
\Psi(\sigma^{-1}x^\prime, \sigma y^\prime):
$
\begin{eqnarray}
&&F[\psi]
=\int d^2{\bf r}^\prime \big(\frac{\hbar^2}{2m}|
{\bf D}^\prime\psi |^2 +\alpha |\psi ({\bf r}^\prime) 
|^2\big) \nonumber \\
&&~+\frac{\beta}{2}\int d^2 {\bf r}^\prime_1 d^2 {\bf r}^\prime_2
|\psi ({\bf r}^\prime_1)|^2 h ({\bf r}^\prime_1
-{\bf r}^\prime_2) |\psi 
({\bf r}^\prime_2)|^2  ,
\label{eqiso}
\end{eqnarray}
where ${\bf D}^\prime=-i{\bf \nabla}^\prime-
(e^*/\hbar c){\bf a} ({\bf r}^\prime)$ and $m=(m_x m_y)^{1/2}$.

Within mean field theory, the structure of the vortex lattice in our model can
be determined in exactly the same way as in Abrikosov's work using the lowest
Landau level approximation (LLL)\cite{abrikosov}. We look for a periodic
solution to the linearized GL equations while restricting the order parameter
to the space spanned by the LLL wavefunctions. In the Landau gauge, ${\bf
a}=(-By^\prime,0)$,  the normalized solution quasi-periodic over the two
periodicity vectors,
${\bf r}^\prime_{\rm I} =l (1,0)$, 
${\bf r}^\prime_{\rm II} =l (\zeta,\eta)$
is given by \cite{eilenberger}
$\psi({\bf r}^\prime)\sim\phi({\bf r}^\prime|0)\equiv
(2\eta)^{1/4}\exp(-\mu^2 y^{\prime 2}/2)\theta_3 (\pi (x^\prime
+i y^\prime)/l|\zeta+i\eta)$
with the theta function $\theta_3$.
The magnetic length
$\mu^{-1}$ is fixed by the flux quantization condition;
$2\pi\mu^{-2}=$(area of unit cell)$=l^2 \eta$. 
A useful representation for $\phi({\bf r}^\prime)$ is \cite{brandt}
\begin{equation}
|\phi({\bf r}^\prime)|^2=\sum_{m,n=-\infty}^\infty 
(-1)^{mn} \exp( -{\bf G}^2/4\mu^2
+i{\bf G}\cdot{\bf r}^\prime ),
\end{equation}
where ${\bf G}=\mu^2 l (\eta m, n-\zeta m)$ 
is the reciprocal lattice vector
corresponding to the periodicity vectors in
the ${\bf r}^\prime$-space. The mean-field free energy 
density is given by
$F_{\rm MF}=-\alpha_H^2 / 2\beta\beta_A,$
where $\alpha_H\equiv\alpha+\hbar e^* B/2mc =0$ determines the
upper critical field $H_{c2}(T)$, and the Abrikosov ratio
$\beta_A$, which 
accounts for the contribution from the quartic term, is given by
\begin{equation}
\beta_A=\sum_{m,n}\exp(-{\bf G}^2/2\mu^2)\; \widetilde{h} ({\bf G}).
\label{betaa}
\end{equation}
In order to find a minimum free energy
configuration, one has to minimize $\beta_A$.

Since the sum in (\ref{betaa}) converges very quickly, it
is not difficult to find a configuration that minimizes $\beta_A$
for given values of $C$, $\varepsilon$ and $\theta_0$.
In the limit where $C\rightarrow 0$ (the low-magnetic field regime),
one recovers the local GL theory, and  
$\beta_A$ attains
the familiar minimum value, $1.159\cdots$ for a triangular lattice in 
the ${\bf r}^\prime$-space.
Also, since rotational invariance exists in the 
${\bf r}^\prime$-space when $C\rightarrow 0$, the triangular lattice has
no preferred orientation with respect to the underlying crystal.
When transformed back to the original ${\bf r}$-space, it results
in a distorted triangular lattice due to the a-b plane
anisotropy. 

In the high field regime where one cannot neglect the
effect of the nonlocal quartic term, the situation is different.
Because of the a-b plane anisotropy carried over to $\widetilde{h}
({\bf k}^\prime)$, the system in the ${\bf r}^\prime$-space
is neither rotationally invariant nor four-fold symmetric. 
Therefore, the vortex lattice
is orientated with respect to the crystal.
In our model, the parameter $\theta_0$ specifies the orientation. 
In general, we obtain an oblique lattice
whose form depends on the values of $C$ and $\varepsilon$. 
In the original ${\bf r}$-space, this lattice is further distorted due to
the a-b plane anisotropy. 
To be more specific,
we parametrize the periodicity vectors,
${\bf r}^\prime_{\rm I}$ and ${\bf r}^\prime_{\rm II}$, in terms of a
centered rectangular lattice for which one can write 
$\zeta=1/2$ and $\eta=(1/2)\tan\varphi^\prime$, where $\varphi^\prime$
is an angle between 
${\bf r}^\prime_{\rm I}$ and ${\bf r}^\prime_{\rm II}$.
We consider a general case where this lattice is rotated by an
angle $\varphi^\prime_0$.  
The resulting lattice can also be regarded as an
oblique lattice with two primitive vectors of equal length and
an angle $2\varphi^\prime$ (or $\pi-2\varphi^\prime$) between them.
For simplicity we focus on the case where $\theta_0=0$.
For given $C$, $\varepsilon$ and $\theta_0=0$, we look for
$\varphi^\prime$ and $\varphi^\prime_0$ that minimizes $\beta_A$. 
For $\theta_0=0$, we find that the minimum free energy 
configuration always corresponds to $\varphi^\prime_0=0$, where one of the
primitive vectors, ${\bf r}^\prime_{\rm I}$ coincides 
with the $x^\prime$-axis. Any other orientation
of the vortex lattice can be obtained using different 
values of $\theta_0$. The angle $\varphi^\prime$ 
that gives the minimum free energy changes continuously 
from $\sim 60^\circ$ corresponding to a distorted
triangular lattice to $\sim 45^\circ$ for a distorted square lattice
as the four-fold symmetric coupling $\varepsilon$ increases 
from 0 toward 1 for fixed $C$. We find that for $C$ greater than
some value $C_c\sim 0.015$, there exists $\varepsilon_c$ which
depends on $C$
such that the vortex lattice remains as a distorted square
lattice for $\varepsilon > \varepsilon_c$. 
A similar behavior to this was obtained in Ref.\onlinecite{new}.
Now, since the original order
parameter $\Psi({\bf r})$ is quasi-periodic with respect to
$l(\sigma^{-1},0)$ and $l(\sigma^{-1}\zeta, \sigma\eta)$
for $\varphi^\prime_0=0$, the oblique lattice in the ${\bf r}$-space 
has an angle $\varphi$, where $\tan\varphi=\sigma^{-2}\tan\varphi^\prime$.
(This relation becomes more complicated if $\varphi^\prime_0\neq 0$.)
To summarise, within
mean-field theory, the structure of the vortex lattice
is mainly determined by the anisotropy parameter
$\sigma$ and the four-fold
symmetric coupling $\varepsilon$, and the  orientation by $\theta_0$. All these
parameters can in principle be fixed by experiments which determine the flux
lattice structure as a function of the magnetic field.

Thermal fluctuations, which are especially important in high-$T_c$
materials, melt the mean-field vortex lattice into a vortex liquid.
For the local theory ($C=0$),
the effect of thermal fluctuations around the mean-field solution
was studied by Eilenberger \cite{eilenberger}
using the orthonormal basis for the LLL wave functions,
$\phi ({\bf r}|{\bf r}_0) =\exp ({\rm i} \mu^2 x y_0)
\phi({\bf r}|0)$, where
${\bf r}_0$ spans one fundamental cell, or
${\bf q} = \mu^2(y_0,-x_0)$ belongs to the first
Brillouin zone. There are two different modes of 
excitation, whose energies are denoted by 
$\epsilon_{\pm}({\bf q})$. In the long wave length limit,
$q \rightarrow 0$, the 
hard mode behaves as $\epsilon_{+}(q)={\rm const}+O(q^2)$ 
while the soft mode takes the form
$\epsilon_{-}(q)= (\alpha_0/2) (q^4/\mu^4)+ O(q^6)$
as $q\rightarrow 0$. The soft mode  corresponds to
an incompressible shear deformation of the vortex
 lattice  \cite{mike2} and  $\alpha_0$ can be identified 
with the shear modulus $c_{66}$ of the
triangular lattice.

When  terms which break the rotational symmetry are present,
the orientation of the
vortex lattice is locked to the underlying crystal.
Therefore we expect that there exists an excitation energy cost
associated with a rigid rotation of the vortex lattice 
against the crystal lattice. We shall calculate this energy when the
coupling to the underlying lattice is 
very weak, {\em i.e.} $C\ll 1$.
Following Ref.~\onlinecite{eilenberger}, we first determine the 
soft mode energy $\epsilon_{-}({\bf q})$ associated with
(\ref{eq1}) to the lowest
order in $C$. (We assume $\sigma=1$ for simplicity.)
After  lengthy but otherwise
straightforward algebra, we 
obtain the following anisotropic expression:
\begin{eqnarray}
&&\epsilon_{-}({\bf q})=\frac{\alpha_0}{2}
\Big[ (1+C \alpha_1)(q^4/\mu^4) \nonumber \\
&&~+C\varepsilon (\beta_1 q^4_{x} + 2\beta_2 q^2_{x} q^2_{y} 
+\beta_3 q^4_{y})/\mu^4 \Big] +O(q^6, C^2) \label{eminus}
\end{eqnarray}
with known {\em numerical} constants $\alpha_1,\beta_1,\beta_2$, 
and $\beta_3$. This corresponds to a general form of the 
effective free energy \cite{kogan}
for the displacement ${\bf u}$, which
should involve the rotation fields, $v_{ij}\equiv
(\partial_i u_j -
\partial_j u_i)/2$ as well as the usual strain fields,
$u_{ij}\equiv
(\partial_i u_j + \partial_j u_i)/2$. In our case
where the vortex lattice is incompressible and four-fold
symmetric, the effective free energy density reduces to
\[
\frac{1}{2}\{ \lambda_1 (\partial_x u_x -\partial_y u_y)^2
+4\lambda_2 u_{xy}^2
+4\omega v_{xy}^2 
+8\xi u_{xy}v_{xy}\}
\]
with four energy constants, 
$\lambda_1, \lambda_2, \omega$, and $\xi$.
For $C=0$, one only has the purely
elastic part ($\omega=\xi=0$) and $\lambda_1=\lambda_2=
c_{66}$. For small $C$, the shear modulus will have a O($C$)
correction, and the rotation modulus $\omega$ and the coupling $\xi$
between the rotation and strain fields will be proportional to 
$C\varepsilon$. From (\ref{eminus}), we obtain using the method of
Ref.~\onlinecite{mike2}  $\lambda_1=c_{66} (1+C\alpha_1 +C\varepsilon 
(2\beta_2-\beta_1-\beta_3)/4)$, $\lambda_2=c_{66}(1+C\alpha_1)$,
$\omega=c_{66}C\varepsilon (\beta_1+\beta_3)/2$, and
$\xi=c_{66}C\varepsilon (\beta_1 -\beta_3)/2$.

A useful quantity in studying the effect of thermal 
fluctuations is
the so-called structure factor $\Delta ({\bf k})$
of the vortex liquid, which
is proportional to the Fourier transform of
the density-density correlation function $\langle|\Psi({\bf r})|^2
|\Psi({\bf r}+{\bf R})|^2\rangle_c$. 
In the low temperature regime, the length scale $l_\perp$
governing the degree of a short-range crystalline order 
perpendicular to the magnetic field in the
vortex liquid becomes very large ($l_\perp \gg l_0$). 
We expect then that a tiny amount of coupling to
the underlying lattice might be able to break the rotational symmetry
of the vortex liquid system even though there is
only a short-range translational order. For example, a ring-like 
pattern expected in a structure factor 
will be broken up into Bragg-like spots even in a liquid state
(these spots will
not be delta-function peaks, {\it i.e.} not true
Bragg peaks). 
This is to be contrasted with the usual explanation of
the appearance of Bragg-like spots in neutron diffraction 
patterns using a phase transition from a vortex liquid 
state to a vortex lattice state. We can estimate the 
angular dispersion $\delta\theta$ of these spots using
the above discussion on the rotation modulus: Since the 
energy associated with a rigid rotation by $\delta\theta$ of 
the crystalline region of area $l^2_\perp$ is given by
$l^2_\perp \omega (\delta\theta)^2$, by equating this to
$k_{\rm B}T$, one finds that 
\begin{equation}
(\delta\theta)^2\sim
k_{\rm B}T/\omega l^2_\perp=(0.012)k_{\rm B}T/c_{66}
l^2_\perp C\varepsilon , \label{delth}
\end{equation} 
where we have used the numerical values 
for $\beta_1$ and $\beta_3$. 

As the temperature is raised, according to (\ref{delth}), one 
needs larger values of $C\varepsilon$ to observe the Bragg-like 
spots. At moderately low temperatures, an approximation
scheme, called the parquet resummation method \cite{ym}
is accessible for the calculation of the structure
factor of the vortex liquid. Using this method, one can
explicitly observe a ring-like pattern in the structure 
factor is broken up into Bragg-like spots as the temperature
is lowered.
For an isotropic system described by 
(\ref{eqiso}), it is straightfoward to apply the parquet resummation
method to calculate the structure 
factor $\Delta^\prime ({\bf k}^\prime)$ in the ${\bf k}^\prime$-space.
The only difference compared to the usual local 
theory is that one starts from 
the bare quartic potential $\widetilde{h}({\bf k}^\prime)$ which 
explicitly breaks the rotational symmetry of the local theory.
For the detailed form of the nonperturbative equations 
one has to solve
numerically for $\Delta^\prime$, and 
the reader is referred to Ref.~\onlinecite{ym} for details.
The structure factor in the original space is given simply by
$\Delta({\bf k})=\Delta^\prime({\bf k}^\prime)$, where
$(k_x^\prime , k_y^\prime)=(\sigma^{-1}k_x, \sigma k_y)$. 

\begin{figure}
\centerline{\epsfxsize=6cm\epsfbox{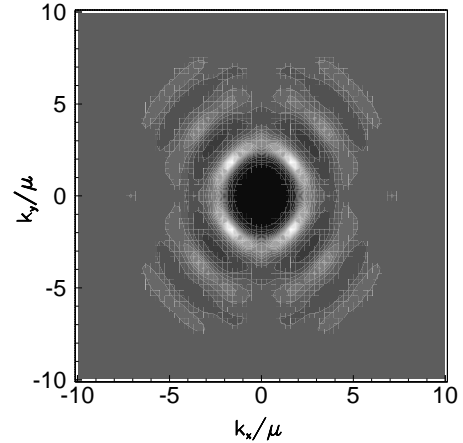}}
\vspace{10pt}
\caption{A contour plot of $\Delta({\bf k})$ at $\alpha_T\simeq -6.3$.
The parameters used are $C=0.01$, $\varepsilon=0.5$, $\theta_0=0$,
and $\sigma^{-2}=1.15$.
}
\label{fig2}
\end{figure}

We have calculated $\Delta ({\bf k})$ for various values of 
the parameters, $C$, $\varepsilon$ and at different temperatures
down to $\alpha_T\sim -6.7$, where
the temperature is represented by the dimensionless quantity,
$\alpha_T\equiv\alpha_H\sqrt{2\pi/\beta\mu^2}$, which goes to
$-\infty$ as one approaches zero temperature.
Fig.~\ref{fig2}
shows a contour plot of $\Delta ({\bf k})$ at $\alpha_T\simeq -6.3$
for the values of $C$ and $\varepsilon$ that correspond to a
moderately strong coupling between the vortex lattice and 
the underlying crystal ($\varepsilon=0.5$).
One can clearly observe four bright spots, which correspond to the
nearest peaks in the structure factor, emerging from a ring-like 
pattern. To observe the next and higher order spots (which will 
give a structure factor a closer resemblance to that
 expected for a triangular lattice)
one has to obtain the structure factor at lower temperatures.
We find in general that,
as the coupling between the vortex lattice
and the underlying crystal gets weaker, 
one has to go to lower temperatures to observe the Bragg-like spots. 
This fact is in qualitative agreement with (\ref{delth})
(although the present numerical calculation is not done in the 
strict weak coupling limit).

In summary, we considered a simple phenomenological model for vortices
in a crystal lattice
using a nonlocal GL theory. As well as explaining
the observed four-fold symmetric vortex lattice structures
within mean-field theory, the
present model suggests that there is 
a possibility of observing Bragg-like spots
within the vortex liquid regime as a consequence of coupling
of the vortices to the underlying crystal.

\end{multicols}
\end{document}